\documentclass[a4paper]{jpconf}
\usepackage[latin1]{inputenc}
\usepackage{amsmath}
\usepackage{amsfonts}
\usepackage{amssymb}
\usepackage{listings}
\usepackage{graphicx}
\usepackage{epstopdf}

\usepackage{amsfonts,amssymb,amscd,amsmath}
\usepackage{amssymb}
\usepackage{amsmath}
\usepackage{graphicx}
\usepackage{graphics}
\usepackage{pifont}
\usepackage{hyperref}

\def\f12{\frac{1}{2}}

\begin{document}
\title{On a Java library to perform S-expansions of Lie algebras}

\author{Carlos Inostroza$^{1}$, Igor Kondrashuk$^{2},$ Nelson Merino$^3$  \\ and Felip Nadal$^4$}
\address{$^1$  Departamento de F\'isica, Universidad de Concepci\'on, Casilla 160-C, Concepci\'on, Chile} 
\address{$^2$  Grupo de Matem\'atica Aplicada, Departamento de Ciencias B\'asicas, Univerdidad del B\'io-B\'io, Campus Fernando May, Casilla 447, Chill\'an, Chile}
\address{$^3$  APC, Universite Paris Diderot, 10, rue Alice Domon et Leonie Duquet, 75205 Paris Cedex 13, France}
\address{$^4$  Instituto de F\'isica Corpuscular (IFIC), Edificio Institutos de Investigaci\'on, \\
                c/ Catedr\'atico Jos\'e Beltr\'an, 2, E-46980 Paterna, Espa\~na}

\ead{nemerino@gmail.com}

\begin{abstract}
The S-expansion method is a generalization of the In\"{o}n\"{u}-Wigner (IW) contraction that allows to study new non-trivial relations between different Lie algebras. 
Basically, this method combines a Lie algebra $\mathcal{G}$ with a finite abelian semigroup $S$ in such a way that a new S-expanded algebra $\mathcal{G}_{S}$ can be defined. 
When the semigroup has a zero-element and/or a specific decomposition, which is said to be resonant with the subspace structure of the original algebra, then it is possible to extract smaller 
algebras from $\mathcal{G}_{S}$ which have interesting properties.
Here we give a brief description of the S-expansion, its applications and the main motivations that lead us to elaborate a Java library, which automatizes this method and allows 
us to represent and to classify all possible S-expansions of a given Lie algebra.
\end{abstract}

\section{Introduction}
As is well-known, the theory of Lie groups and algebras plays an essential role in Physics: it represents the mathematical tool allowing to describe the continues symmetries of a physical system and, via the Noether theorem, this is directly connected with the corresponding conservation laws of the system. Since the fifties, different mechanisms allowing to establish non-trivial relations between Lie algebras appeared and were very useful to understand interrelations between different physical theories. The original idea was introduced in Ref. \cite{Segal}, where it was pointed out that if two physical theories are related by means of a limit process (like Newtonian mechanics and special relativity which are related by the limit where the speed of light $c$ goes to infinite), then the corresponding symmetry groups (the Galilean and Poincar\'e groups, in the example) under which those theories are invariant should also be related by means of a similar limit process. This process was formally introduced in Ref. \cite{IW,IW2} and it is known nowdays as In\"{o}n\"{u}-Wigner (IW) contraction. 

Many generalized contractions were introduced in the literature during the last decades. In particular, the Weimar-Woods (WW) contraction \cite{WW2,WW,WW3} is one of the most general realization of this method. Starting with an algebra which have a certain subspace structure, the WW contraction performs a suitable rescaling with a real parameter on the generators of each subspace. Then, a special limit for that parameter leads to the contracted algebra, which has the same dimension than the original one, but very different properties\footnote{For example, if the original algebra is semisimple, then the contracted algebra is usually non-semisimple.}.

Another peculiar generalization of the contraction, known as \textit{expansion} method, was parallelly introduced in the context of string theory \cite{hs} and supergravity \cite{aipv1}. This procedure not only is able to reproduce the WW contractions when the dimension is preserved in the process, but also may lead to expanded algebras whose dimension is higher than the original one. The main difference with the contraction method is that, using the dual description of a Lie algebra in terms of Maurer-Cartan (MC) forms, the rescaling is performed on some coordinates of the Lie group manifold and not on the generators of the Lie algebra. 
%Between other applications, this procedure has been used to study a relation between five-dimensional General Relativity and  Chern-Simons Gravity \cite{Edelstein:2006se}.
 
Here we focus on an even more general procedure called \textit{S-expansion}\footnote{The dual formulation of the S-expansion in terms of MC forms was performed in \cite{irs2}, while in Refs. \cite{Caroca:2010ax,Caroca:2010kr,Caroca:2011zz} the method was also extended to other mathematical structures, like the case of higher order Lie algebras and infinite dimensional loop algebras.} \cite{irs}, which combines the structure constants of the original algebra with the inner multiplication law of an abelian finite semigroup of order $n$ to define a new \textit{S-expanded algebra} $\mathcal{G}_{S}$. 
When the semigroup have a zero-element and/or a specific decomposition, which is said to be resonant with the subspace structure of the original algebra, then it is possible to extract smaller algebras, called \textit{resonant subalgebras} and \textit{reduced algebras}.
In particular, the previous expansion method \cite{hs,aipv1} can be reproduced as a $0_{S}$-reduction of the resonant subalgebra for an expansion with a special family of semigroups denoted by $S_{E}^{\left(  N\right)}$.

During the last decade, many applications using the S-expansion has been performed \cite{Izaurieta:2006aj,Izaurieta:2009hz,Quinzacara:2012zz,Concha:2013uhq,Concha:2014zsa,Quinzacara:2013uua,Crisostomo:2014hia,Crisostomo:2016how}, 
mainly in the context of modified theories of gravity. 
At the begining, they considered only the family of semigroups $S_{E}^{(N)}$, which lead to the definition of the so called $\mathfrak{B}_{N}$ algebras.
The use of other abelian semigroups to perform S-expansions of Lie algebras was first considered in Refs. \cite{Caroca:2011qs}, where it was pointed out that possible new applications could be performed 
if the following question is analyzed: \textit{given two Lie algebras, is it possible to find a suitable semigroup that relates them by means of an S-expansion?} 
Of course, the answer depends on the specific algebras that we want to connect and involves to consider all possible finite abelian semigroups in each order\footnote{An example, it was considered the 2 and 3-dimensional 
that acts transitively on 2 and 3-dimensional spaces, i.e., the two 2-dimensional algebras $[X_{1},X_{2}]=0$, $[X_{1},X_{2}]=X_{1}$ and the ten 3-dimensional algebras classified by Bianchi \cite{bian}. It was shown, 
that only four Bianchi algebras can be obtained as S-expansions of the the 2-dimensional algebras if one uses some specific semigroups of order 4, which do not belong in general to the $S_{E}^{(N)}$ family. 
Thus, this result cannot be obtained neither by a contraction nor by the expansion method \cite{aipv1}.
The procedure used to construct by hand the multiplication table of those semigroups also made clear that, if a given problem involves the use of semigroups of higher order, then the use of computer programs is needed.} 
(for a brief review about that classification see the presentation \cite{PresCI}, by C. Inostroza).  
In the same line, a study of the general properties $S$-expansion method (in the context of the classification of Lie algebras) was made in Ref. \cite{Andrianopoli:2013ooa} and revealed that some properties 
of the semigroup allow to determine if a given property of the original algebra (like semisimplicity, compactness, etc) will be preserved or not under the expansion process. 
These results were shown to be useful as criteria to answer if two given algebras can be S-related.
In particular, semigroups preserving semisimplicity were identified.

The results given in Refs. \cite{Caroca:2011qs} and \cite{Andrianopoli:2013ooa} (and in particular the use of semigroups preserving semisimplicity) were used in Ref. \cite{Diaz:2012zza} 
to show that the semisimple version of the so called Maxwell algebra (introduced in \cite{Soroka:2006aj}) can be obtained as an expansion of the AdS algebra. Later, 
this result was generalized in Refs. \cite{Salgado:2014qqa,Concha:2016hbt} to new families of semigroups generating algebras denoted by
$\mathfrak{C}_{N}$ and $\mathfrak{D}_{N}$ which have been useful to construct new (super)gravity models 
\cite{Salgado:2014jka,Fierro:2014lka,Concha:2014vka,Concha:2014xfa,Concha:2014tca,Gonzalez:2014tta,Concha:2015tla,Concha:2015woa,Concha:2016kdz,Gonzalez:2016xwo,Durka:2016eun,Concha:2016tms,Ipinza:2016con,Concha:2016zdb,Penafiel:2016ufo,Artebani:2016gwh,Ipinza:2016bfc}. 
These several new applications show the importance of considering also semigroups outside of the $S_{E}^{(N)}$ family. 
Motivated by this, we have constructed a Java library \cite{webJava,Inostroza:2017ezc} that automatizes the S-expansion procedure and that is aimed to create a general picture about all possible S-relations between Lie algebras.

\section{Brief review of the S-expansion method}

Consider a Lie algebra $\mathcal{G}$ with generators $\{X_{i}\}$ and Lie
product $\left[  X_{i},X_{j}\right]  =C_{ij}^{k}X_{k}$ where $C_{ij}^{k}$ are the structure constants. Consider also a finite
abelian semigroup $S=\{\lambda_{\alpha},\alpha=1,\ldots,n\}$.
An informal way (but useful for our purposes) of expressing the semigroup multiplication law
is by means of quantities called \textit{selectors}\footnote{In particular, the selectors provide a matrix representation for the semigroup
$\lambda_{\alpha}\rightarrow\left(  \lambda_{\alpha}\right)  _{\ \beta}^{\rho}=K_{\alpha\beta}^{\rho}$, which is used in our library to represent a given semigroup.}, denoted by $K_{\alpha\beta}^{\kappa}$ 
and defined by the relation $\lambda_{\alpha}\cdot\lambda_{\beta} =\lambda_{\gamma\left(  \alpha,\beta\right)  }=K_{\alpha\beta}^{\rho}\lambda_{\rho}$, 
where $K_{\alpha\beta}^{\rho}=1$ if $\rho=\gamma\left(  \alpha,\beta\right)  $ and $K_{\alpha\beta}^{\rho}=0$ if $\rho\neq\gamma\left(  \alpha,\beta\right)  \,$. Then, the S-expansion consists of the following steps.
\medskip

\textbf{Step I: Constructing S-expanded algebra.} As shown in \cite{irs}, the set $\mathcal{G}_{S}$\ $\mathcal{=}$\ $S\otimes\mathcal{G}$ (with $\otimes$ being the Kronecker product between
the matrix representation of $S$ and $\mathcal{G}$) is also Lie algebra, which
is called \textit{expanded algebra}, if the basis elements are defined as
$X_{\left(  i,\alpha\right)  }\equiv\lambda_{\alpha}\otimes X_{i}$ and Lie
product by $\left[  X_{\left(  i,\alpha\right)  },X_{\left(  j,\beta\right)  }\right]
\equiv\lambda_{\alpha}\cdot\lambda_{\beta}\otimes\left[  X_{i},X_{j}\right]
=C_{\left(  i,\alpha\right)  \left(  j,\beta\right)  }^{\left(  k,\gamma
\right)  }X_{\left(  k,\gamma\right)  }\label{z3}$.
The structure constants of the expanded algebra $\mathcal{G}_{S}$ are
fully determined by the selectors and the structure constants of the original
Lie algebra $\mathcal{G}$, i.e., $C_{\left(  i,\alpha\right)  \left(
j,\beta\right)  }^{\left(  k,\gamma\right)  }=K_{\alpha\beta}^{\gamma}%
C_{ij}^{k}\,$. 
%In the proof it can be seen that the commutativity property of the semigroup is crucial for the Jacobi identity to be satisfied in $\mathcal{G}_{S}$. 
\medskip

\textbf{Step II: Extraction of the resonant subalgebra.} Consider the case where the original algebra has the
subspace decomposition $\mathcal{G}=V_{0}\oplus V_{1}$ with the following
structure
\begin{equation}
\left[  V_{0},V_{0}\right]  \subset V_{0}\,,\ \ \ \left[  V_{0},V_{1}\right]
\subset V_{1}\,,\ \ \ \left[  V_{1},V_{1}\right]  \subset V_{0}\label{r1}%
\end{equation}
Suppose also that a given semigroup has a decomposition in subsets $S=S_{0}\cup S_{1}$ satisfying
\begin{equation}
S_{0}\cdot S_{0}\in S_{0}\,,\ \ \ S_{0}\cdot S_{1}\in S_{1}\,,\ \ \ S_{1}\cdot
S_{1}\in S_{0}\label{r2}%
\end{equation}
which is called \textit{resonant condition}, because of the similarity with the
subspace structure (\ref{r1}) of the algebra\footnote{As it can be seen in \cite{irs}, one can deal with
algebras and semigroups having decompositions which are more general, but in the first version of our library we only consider the case given
by (\ref{r1}) and (\ref{r2}).}. Then, as shown in \cite{irs},
the set $\mathcal{G}_{S,R}=\left(  S_{0}\otimes V_{0}\right)  \oplus\left(S_{1}\otimes V_{1}\right)$ is a subalgebra of the expanded algebra $\mathcal{G}_{S}$\ $\mathcal{=}%
$\ $S\otimes\mathcal{G}$.
\medskip

\textbf{Step III: Extraction of the }$0_{S}$-\textbf{reduced algebra.} If semigroup contains an element $0_{S}$ satisfying $\lambda_{\alpha}%
\cdot0_{S}=0_{S}$ for any element $\lambda_{\alpha}\in S$, then this element
is called a zero element. In that case, the sector $0_{S}\otimes\mathcal{G}$
can be removed from the expanded algebra in such a way that what is left is
also a Lie algebra, called the $0_{S}$\textit{-reduced algebra}. Remarkably, the reduced algebra is not necessarily a subalgebra of the expanded algebra.% $\mathcal{G}_{S}$. 

\section{Need of automizing the procedure}
First we notice that the steps II and III are independent, but can also be applied simultaneously. This means that, depending on the semigroup that is going to be used in the expansion, the following algebras can be obtained: 
\begin{enumerate}
	\item the expanded algebra $\mathcal{G}_{S} = S \otimes \mathcal{G}$,
	\item the resonant subalgebra $\mathcal{G}_{S,R}=\left(  S_{0}\otimes V_{0}\right)  \oplus\left(S_{1}\otimes V_{1}\right)$, if the semigroup have at least one resonant decomposition,
	\item the  $0_{S}$-reduced algebra if the semigroup have a zero element  $0_{S}$,
	\item the  $0_{S}$-reduction of the resonant subalgebra, if the semigroup has simultaneously a zero element and at least one resonant decomposition.
\end{enumerate}

In order to study all possible S-expansions (i-iv) of a given algebra, one should consider the full set of abelian semigroups. As it will be reviewed in the presentation by C. Inostroza \cite{PresCI}, 
the problem of enumerating the all non-isomorphic finite semigroups of a certain order is a non-trivial problem because the number of semigroups increases very quickly with the order of the semigroup. 
Thus, this task can be performed only up to a certain order and, in particular, we have used the program \textit{gen.f} of Ref. \cite{Hildebrant} to generate the files \textit{sem.2}, \textit{sem.3}, 
\textit{sem.4}, \textit{sem.5} and \textit{sem.6} which contain all the non isomorphic semigroups up to order 6. Those files can be used as input data for many of the programs that compose our library 
\cite{webJava,Inostroza:2017ezc}.

\begin{figure}[th]
\centering
\includegraphics[scale=0.5]{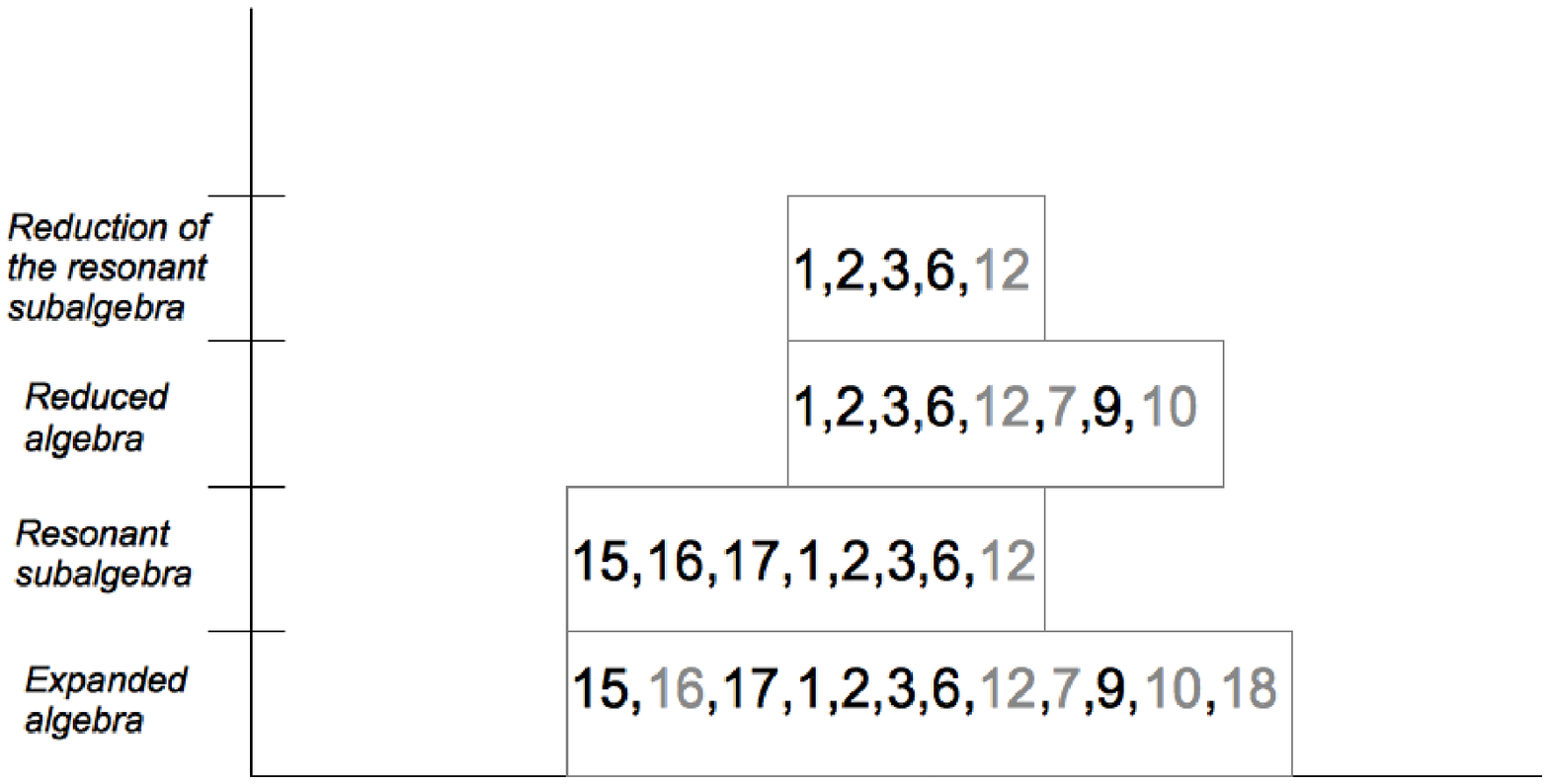}
\caption{Representing different kind of expansions with semigroups of order $3$.}%
\label{fig:fig0}
\end{figure}

For example, in the order 3 there are 18 non-isomorphic semigroups denoted by $S_{(3)}^{a} $ with $a=1,...18$, from which only 12 of them 
are abelian\footnote{The non-isomorphic abelian semigroups of order 3 are: $$\begin{tabular}
[c]{l|lll}%
$S_{\left(  3\right)  }^{1}$ & $\lambda_{1}$ & $\lambda_{2}$ & $\lambda_{3}%
$\\\hline
$\lambda_{1}$ & $\lambda_{1}$ & $\lambda_{1}$ & $\lambda_{1}$\\
$\lambda_{2}$ & $\lambda_{1}$ & $\lambda_{1}$ & $\lambda_{1}$\\
$\lambda_{3}$ & $\lambda_{1}$ & $\lambda_{1}$ & $\lambda_{1}$%
\end{tabular}
\ \ \text{, }%
\begin{tabular}
[c]{l|lll}%
$S_{\left(  3\right)  }^{2}$ & $\lambda_{1}$ & $\lambda_{2}$ & $\lambda_{3}%
$\\\hline
$\lambda_{1}$ & $\lambda_{1}$ & $\lambda_{1}$ & $\lambda_{1}$\\
$\lambda_{2}$ & $\lambda_{1}$ & $\lambda_{1}$ & $\lambda_{1}$\\
$\lambda_{3}$ & $\lambda_{1}$ & $\lambda_{1}$ & $\lambda_{2}$%
\end{tabular}
\ \ \text{, }%
\begin{tabular}
[c]{l|lll}%
$S_{\left(  3\right)  }^{3}$ & $\lambda_{1}$ & $\lambda_{2}$ & $\lambda_{3}%
$\\\hline
$\lambda_{1}$ & $\lambda_{1}$ & $\lambda_{1}$ & $\lambda_{1}$\\
$\lambda_{2}$ & $\lambda_{1}$ & $\lambda_{1}$ & $\lambda_{1}$\\
$\lambda_{3}$ & $\lambda_{1}$ & $\lambda_{1}$ & $\lambda_{3}$%
\end{tabular}
\ \ \text{, }%
\begin{tabular}
[c]{l|lll}%
$S_{\left(  3\right)  }^{6}$ & $\lambda_{1}$ & $\lambda_{2}$ & $\lambda_{3}%
$\\\hline
$\lambda_{1}$ & $\lambda_{1}$ & $\lambda_{1}$ & $\lambda_{1}$\\
$\lambda_{2}$ & $\lambda_{1}$ & $\lambda_{1}$ & $\lambda_{2}$\\
$\lambda_{3}$ & $\lambda_{1}$ & $\lambda_{2}$ & $\lambda_{3}$%
\end{tabular}
\ \ \text{, }%
$$
$$
\begin{tabular}
[c]{l|lll}%
$S_{\left(  3\right)  }^{7}$ & $\lambda_{1}$ & $\lambda_{2}$ & $\lambda_{3}%
$\\\hline
$\lambda_{1}$ & $\lambda_{1}$ & $\lambda_{1}$ & $\lambda_{1}$\\
$\lambda_{2}$ & $\lambda_{1}$ & $\lambda_{2}$ & $\lambda_{1}$\\
$\lambda_{3}$ & $\lambda_{1}$ & $\lambda_{1}$ & $\lambda_{3}$%
\end{tabular}
\ \text{, }%
\begin{tabular}
[c]{l|lll}%
$S_{\left(  3\right)  }^{9}$ & $\lambda_{1}$ & $\lambda_{2}$ & $\lambda_{3}%
$\\\hline
$\lambda_{1}$ & $\lambda_{1}$ & $\lambda_{1}$ & $\lambda_{1}$\\
$\lambda_{2}$ & $\lambda_{1}$ & $\lambda_{2}$ & $\lambda_{2}$\\
$\lambda_{3}$ & $\lambda_{1}$ & $\lambda_{2}$ & $\lambda_{2}$%
\end{tabular}
\ \text{, }%
\begin{tabular}
[c]{l|lll}%
$S_{\left(  3\right)  }^{10}$ & $\lambda_{1}$ & $\lambda_{2}$ & $\lambda_{3}%
$\\\hline
$\lambda_{1}$ & $\lambda_{1}$ & $\lambda_{1}$ & $\lambda_{1}$\\
$\lambda_{2}$ & $\lambda_{1}$ & $\lambda_{2}$ & $\lambda_{2}$\\
$\lambda_{3}$ & $\lambda_{1}$ & $\lambda_{2}$ & $\lambda_{3}$%
\end{tabular}
\ \text{, }%
\begin{tabular}
[c]{l|lll}%
$S_{\left(  3\right)  }^{12}$ & $\lambda_{1}$ & $\lambda_{2}$ & $\lambda_{3}%
$\\\hline
$\lambda_{1}$ & $\lambda_{1}$ & $\lambda_{1}$ & $\lambda_{1}$\\
$\lambda_{2}$ & $\lambda_{1}$ & $\lambda_{2}$ & $\lambda_{3}$\\
$\lambda_{3}$ & $\lambda_{1}$ & $\lambda_{3}$ & $\lambda_{2}$%
\end{tabular}
\ \text{, }%
$$
$$
\begin{tabular}
[c]{l|lll}%
$S_{\left(  3\right)  }^{15}$ & $\lambda_{1}$ & $\lambda_{2}$ & $\lambda_{3}%
$\\\hline
$\lambda_{1}$ & $\lambda_{1}$ & $\lambda_{1}$ & $\lambda_{3}$\\
$\lambda_{2}$ & $\lambda_{1}$ & $\lambda_{1}$ & $\lambda_{3}$\\
$\lambda_{3}$ & $\lambda_{3}$ & $\lambda_{3}$ & $\lambda_{1}$%
\end{tabular}
\ \text{, }%
\begin{tabular}
[c]{l|lll}%
$S_{\left(  3\right)  }^{16}$ & $\lambda_{1}$ & $\lambda_{2}$ & $\lambda_{3}%
$\\\hline
$\lambda_{1}$ & $\lambda_{1}$ & $\lambda_{1}$ & $\lambda_{3}$\\
$\lambda_{2}$ & $\lambda_{1}$ & $\lambda_{2}$ & $\lambda_{3}$\\
$\lambda_{3}$ & $\lambda_{3}$ & $\lambda_{3}$ & $\lambda_{1}$%
\end{tabular}
\ \text{, }%
\begin{tabular}
[c]{l|lll}%
$S_{\left(  3\right)  }^{17}$ & $\lambda_{1}$ & $\lambda_{2}$ & $\lambda_{3}%
$\\\hline
$\lambda_{1}$ & $\lambda_{1}$ & $\lambda_{2}$ & $\lambda_{2}$\\
$\lambda_{2}$ & $\lambda_{2}$ & $\lambda_{1}$ & $\lambda_{1}$\\
$\lambda_{3}$ & $\lambda_{2}$ & $\lambda_{1}$ & $\lambda_{1}$%
\end{tabular}
\ \text{, }%
\begin{tabular}
[c]{l|lll}%
$S_{\left(  3\right)  }^{18}$ & $\lambda_{1}$ & $\lambda_{2}$ & $\lambda_{3}%
$\\\hline
$\lambda_{1}$ & $\lambda_{1}$ & $\lambda_{2}$ & $\lambda_{3}$\\
$\lambda_{2}$ & $\lambda_{2}$ & $\lambda_{3}$ & $\lambda_{1}$\\
$\lambda_{3}$ & $\lambda_{3}$ & $\lambda_{1}$ & $\lambda_{2}$%
\end{tabular}
\ \text{, }$$}.
Thus, the different type of expansions that can be done with them are represented in the figure \ref{fig:fig0} where, for reasons of space, we use only the label `$a$' to name the different semigroups in the horizontal axis, while in the vertical axis we represent the different kinds of expansions that can be performed with them. With a \textit{gray number} it has also been identified the expansions that preserve the semisimplicity.

%For example, in the order 3 there are 18 non-isomorphic semigroups denoted by $S_{(3)}^{a} $ with $a=1,...18$. For reasons of space, in the figure \ref{fig:fig0} we use only the label to name the different `$a$'. The horizontal axis represents the set of semigroups used in some specific expansion while the vertical axis represents the different kinds of expansions that can be performed.
%Thus, they allow to perform the expansions of the type (i).
%To identify all possible resonant decompositions of all non-isomorphic semigroups of a given order is neither a trivial task, even for the simplest case described by (\ref{r1}) and (\ref{r2}) which we have considered in the first version of our library. Thus, the first version of the library we have constructed 
To identify all possible resonant decompositions of all non-isomorphic semigroups in higher orders is not a trivial task. For this reason, we have constructed a Java library \cite{webJava,Inostroza:2017ezc} which allows us to study all possible S-expansions of the type (i-iv) using the mentioned lists \textit{sem.n}.

\section{On the maximal order of semigroups which the Java Library can perform}

In some cases it is useful to have the full lists of non-isomorphic semigroups up to certain order. However, there is an intrinsic computational limitation to construct these lists for order 8 and higher. 
Indeed, the number of non-isomorphic semigroups in those orders has been evaluated only by using indirect techniques. For lower orders, instead,  the construction of these tables is possible. 
For example, in Ref. \cite{Hildebrant}
a fortran program  gen.f was proposed, which, according to what is claimed in this reference, is able to generate these lists up to order 7. However, after running that fortran program  we were able to obtain 
those lists only up to order 6. These lists are used as inputs for our library \cite{webJava} although its methods are not restricted to the order 6. Indeed, the methods of our library also allow us to work 
with semigroup of higher orders. 
The only issue is that we do not have the full list of non-isomorphic tables for those higher orders. For example, our methods are able to read a multiplication table of order 20 and show in few seconds if 
that table is associative, commutative, to identify if there is a zero element and also allow to check if a given decomposition is resonant. 
The only task that involves more computations and time is the method {\it findAllResonances} that finds out all possible resonant decompositions of a given semigroup. 
In that case, one can obtain the answer in few minutes for semigroups of order 8, during about one hour for semigroups of order 9, 
and in our case we were unable to obtain an answer over 3 days for semigroups of order 10. All the computations above were made in usual laptop computers.

\section{Conclusions}

We have presented a brief description of the S-expansion method, its applications and the motivations to automatize this procedure. 
We have made this with a set of methods organized in the form of Java Library \cite{webJava,Inostroza:2017ezc} which is able not only to represent and perform basic operations with semigroups and 
Lie algebras but also to perform operations that are intrinsic of the S-expansion method, such as the characterization of resonances and the corresponding representation of the resonant subalgebra and the reduced algebras. 
The main features of this library will be described in the related talk by C. Inostroza \cite{PresCI}.

%As mentioned in \cite{Nesterenko2012} it it would be interesting to analyze if the S-expansion could help to fit to the classification of solvable Lie algebras of a fixed dimension using S-expansions of semisimple Lie algebras of the same dimension. Thus, apart from new physical applications, our library might also be useful to analyze that problem.

\ack{We thank Andrei Kataev for his invitation to present this work at ACAT 2017. C.I. was supported by a Mecesup PhD grant and the T\'ermino de tesis grant from CONICYT (Chile). C.I. is very grateful to Local 
Organizing Committee and in particular to Gordon Watts for the financing his participation in ACAT.  
I.K. was supported by was supported by Fondecyt (Chile) grant 1050512 and by DIUBB (Chile) Grant Nos. 102609 and GI 153209/C. 
NM is supported by a Becas-Chile postdoctoral grant.}


\section*{References}

\begin{thebibliography}{99}

\bibitem {Segal} Segal I E  1951  A class of operator algebras which are determined by groups  
{\it Duke Math. J.} {\bf 18} 221-65 

\bibitem {IW} In\"{o}n\"{u} E and Wigner E P 1953 On the contraction of groups and their representations 
{\it Proc. Nat. Acad. Sci. USA} {\bf 39} 510-24 

\bibitem {IW2} In\"{o}n\"{u} E  1964 Contractions of Lie groups and their representations {\it  Group theoretical concepts in elementary particle physics}, ed G\"{u}rsey F, (New York: Gordon and Breach) pp 391-402 

\bibitem {WW2} Weimar-Woods E 1991 Contractions of Lie algebras: generalized Inonu-Wigner contractions versus graded contractions {\it J. Math. Phys.} {\bf 32}  2028 

\bibitem {WW} Weimar-Woods E 1995 Contractions of Lie algebras: generalized Inonu-Wigner contractions versus graded contractions {\it J. Math. Phys.} {\bf 36}  4519-48 


\bibitem {WW3} Weimar-Woods E 2000 Contractions, generalized In\"{o}n\"{u} and Wigner contractions and deformations of finite-dimensional Lie algebras 
{\it Rev. Math. Phys.} {\bf 12} 1505-29 

\bibitem {hs} Hatsuda M and Sakaguchi M 2003 Wess-Zumino term for the AdS superstring and generalized Inonu-Wigner  contraction 
{\it Prog. Theor. Phys.} {\bf 109} 853  ({\it Preprint} hep-th/0106114 ) 


\bibitem {aipv1} de Azcarraga J A, Izquierdo J M, Picon M  and Varela O 2003 Generating Lie and gauge free differential (super)algebras by expanding Maurer-Cartan forms and Chern-Simons supergravity  
{\it Nucl. Phys.} B {\bf 662} 185  ({\it Preprint} hep-th/0212347)


%\bibitem {Edelstein:2006se} J.~D.~Edelstein, M.~Hassaine, R.~Troncoso and J.~Zanelli,  ``Lie-algebra expansions, Chern-Simons theories and the Einstein-Hilbert Lagrangian,''  
%Phys.\ Lett.\ B \textbf{640}, 278 (2006) [hep-th/0605174].
%%CITATION = HEP-TH/0605174;%%

\bibitem {irs} Izaurieta F, Rodriguez E and Salgado P 2006  Expanding Lie (super)algebras through Abelian semigroups  
{\it J. Math. Phys.} {\bf 47} 123512  ({\it Preprint} hep-th/0606215)


\bibitem {irs2} Izaurieta F,  Rodriguez E, Perez A and Salgado P 2009  Dual Formulation of the Lie Algebra S-expansion Procedure 
{\it J. Math. Phys.} {\bf 50} 073511  ({\it Preprint} arXiv:0903.4712 [hep-th])


\bibitem{Caroca:2010ax}  Caroca R, Merino N and Salgado P 2009 S-Expansion of Higher-Order Lie Algebras 
{\it J.\ Math.\ Phys.}  {\bf 50} 013503  ({\it Preprint} arXiv:1004.5213 [math-ph])


\bibitem{Caroca:2010kr} Caroca R, Merino N, Perez A and Salgado P 2009 Generating Higher-Order Lie Algebras by Expanding Maurer Cartan Forms
{\it J.\ Math.\ Phys.}  {\bf 50} 123527  ({\it Preprint} arXiv:1004.5503 [hep-th])


\bibitem{Caroca:2011zz} Caroca R, Merino N, Salgado P and Valdivia O 2011 Generating infinite-dimensional algebras from loop algebras by expanding Maurer-Cartan forms
{\it J.\ Math.\ Phys.}  {\bf 52} 043519 


\bibitem{Izaurieta:2006aj} Izaurieta F, Rodriguez E and Salgado P 2008 Eleven-dimensional gauge theory for the M algebra as an Abelian semigroup expansion of osp(32|1)
{\it Eur.\ Phys.\ J.} C {\bf 54}  675 ({\it Preprint}  hep-th/0606225) 


\bibitem {Izaurieta:2009hz} Izaurieta F, Rodriguez E, Minning P, Salgado P and Perez A 2009 Standard General Relativity from Chern-Simons Gravity
{\it Phys.\ Lett.} B {\bf 678}  213  ({\it Preprint} arXiv:0905.2187 [hep-th] )


\bibitem{Quinzacara:2012zz} Quinzacara C A C and Salgado P 2012 Black hole for the Einstein-Chern-Simons gravity 
{\it Phys.\ Rev.} D {\bf 85}  124026  ({\it Preprint} arXiv:1401.1797 [gr-qc] )


\bibitem {Concha:2013uhq} Concha P K,  Pe\~nafiel D M, Rodr\'iguez  E K and Salgado P 2013 Even-dimensional General Relativity from Born-Infeld gravity 
{\it Phys.\ Lett.}  B  {\bf 725} 419   ({\it Preprint} arXiv:1309.0062 [hep-th])  


\bibitem{Concha:2014zsa} Concha P K, Pe\~nafiel D M, Rodr\'iguez E K and Salgado P 2015 Generalized Poincar\'e algebras and Lovelock-Cartan gravity theory 
{\it Phys.\ Lett.} B {\bf 742}  310   ({\it Preprint} arXiv:1405.7078 [hep-th])
	
	
\bibitem{Quinzacara:2013uua} Quinzacara C A C and Salgado P 2013 Stellar equilibrium in Einstein-Chern-Simons gravity 
{\it Eur.\ Phys.\ J.} C {\bf 73} 2479 


\bibitem{Crisostomo:2014hia} Cris\'ostomo J, G\'omez F, Salgado P, Quinzacara C, Cataldo M and del Campo S 2014 Accelerated FRW Solutions in Chern-Simons Gravity 
{\it Eur.\ Phys.\ J.} C {\bf 74} 3087    ({\it Preprint} arXiv:1401.2128 [gr-qc]) 


\bibitem{Crisostomo:2016how} Cris\'ostomo J, G\'omez F, Quinzacara C and Salgado P 2016 Static solutions in Einstein-Chern-Simons gravity 
{\it J. Cosmol. Astropart. Phys.} JCAP 1606 (2016) no.06, 049  ({\it Preprint} arXiv:1601.06592 [gr-qc]) 



\bibitem {Caroca:2011qs}  Caroca R, Kondrashuk I, Merino N  and Nadal F 2013 Bianchi spaces and their three-dimensional isometries as S-expansions of two-dimensional isometries
{\it J.\ Phys.} A {\bf 46} 225201   ({\it Preprint} arXiv:1104.3541 [math-ph])


\bibitem {bian} Bianchi L  1898 Sugli spazi a tre dimensioni che ammettono un gruppo continuo di movimenti  
{\it Memorie \ di Matematica e di Fisica della Societa Italiana delle Scienze} Serie Terza  {\bf Tomo XI} 267-352 


\bibitem {PresCI} Inostroza C, Kondrashuk I, Merino N and Nadal F 2017 On the algorithm to find S-related Lie algebras {\it Merged talk by C Inostroza at ACAT 2017} 


\bibitem {Andrianopoli:2013ooa} Andrianopoli L, Merino N, Nadal F and Trigiante M 2013 General properties of the expansion methods of Lie algebras 
{\it J.\ Phys.} A {\bf 46} 365204    ({\it Preprint} arXiv:1308.4832 [gr-qc])


\bibitem {Diaz:2012zza} D\'{\i}az J D, Fierro O, Izaurieta F, Merino N, Rodr\'{\i}guez E, Salgado P and Valdivia O  2012 A generalized action for (2 + 1)-dimensional Chern-Simons gravity
{\it J.\ Phys.} A {\bf 45} 255207   ({\it Preprint} arXiv:1311.2215 [gr-qc]) 


\bibitem {Soroka:2006aj} Soroka  D V and Soroka V A 2009  Semi-simple extension of the (super)Poincare algebra 
{\it Adv.\ High Energy Phys.}  234147   ({\it Preprint} hep-th/0605251) 


\bibitem {Salgado:2014qqa} Salgado P and  Salgado S 2014 $\mathfrak{so}(D-1,1)\otimes\mathfrak{so}(D-1,2)$ algebras and gravity  
{\it Phys.\ Lett.} B {\bf 728}  5







\bibitem{Concha:2016hbt} Concha P K, Durka R, Merino N and Rodr\'iguez E K 2016 New family of Maxwell like algebras 
{\it Phys.\ Lett.} B {\bf 759}  507   ({\it Preprint} arXiv:1601.06443 [hep-th])



\bibitem{Salgado:2014jka}  Salgado P, Szabo R J and Valdivia O  2014 Topological gravity and transgression holography
{\it Phys.\ Rev.}  D  {\bf 89} 084077   ({\it Preprint} arXiv:1401.3653 [hep-th])



\bibitem{Fierro:2014lka} Fierro O, Izaurieta F, Salgado P  and Valdivia O  2014 (2+1)-dimensional supergravity invariant under the AdS-Lorentz superalgebra
({\it Preprint} arXiv:1401.3697 [hep-th])



\bibitem {Concha:2014vka} Concha P K, Pe\~nafiel D M,  Rodr\'iguez E K and P.~Salgado P 2014 Chern-Simons and Born-Infeld gravity theories and Maxwell algebras type
{\it Eur.\ Phys.\ J.} C {\bf 74} 2741   ({\it Preprint} arXiv:1402.0023 [hep-th])



\bibitem{Concha:2014xfa}  Concha P K  and  Rodr\'iguez E K 2014 Maxwell Superalgebras and Abelian Semigroup Expansion 
{\it Nucl.\ Phys.} B {\bf 886} 1128   ({\it Preprint} arXiv:1405.1334 [hep-th]) 



\bibitem{Concha:2014tca} Concha P K  and Rodr\'iguez E K  2014 N = 1 Supergravity and Maxwell superalgebras 
{\it J. High Energy Phys.} JHEP09(2014)090   ({\it Preprint} arXiv:1407.4635 [hep-th])  
	

\bibitem{Gonzalez:2014tta} Gonz\'{a}lez N, Salgado P, Rubio G and Salgado S 2014 Einstein-Hilbert action with cosmological term from Chern-Simons gravity 
{\it J. Geom.  Phys.}  {\bf 86}  339 


\bibitem{Concha:2015tla} Concha P K, Rodr\'iguez E K and  Salgado P 2015 Generalized supersymmetric cosmological term in $N=1$ Supergravity 
{\it J. High Energy Phys.}  JHEP08(2015)009    ({\it Preprint}  arXiv:1504.01898 [hep-th]) 


\bibitem{Concha:2015woa} Concha P K, Fierro O, Rodr\'iguez E K and Salgado P 2015 Chern-Simons supergravity in $D=3$ and Maxwell superalgebra 
{\it Phys.\ Lett.} B {\bf 750} 117   ({\it Preprint}  arXiv:1507.02335 [hep-th])


\bibitem{Concha:2016kdz} Concha P K, Durka R, Inostroza C, Merino N  and Rodr\'iguez E K 2016 Pure Lovelock gravity and Chern-Simons theory 
{\it Phys.\ Rev.} D {\bf 94} 024055   ({\it Preprint}  arXiv:1603.09424 [hep-th] )


\bibitem{Gonzalez:2016xwo} Gonz\'{a}lez N, Rubio G, Salgado P and Salgado S 2016 Generalized Galilean algebras and Newtonian gravity
{\it Phys.\ Lett.} B {\bf 755} 433  ({\it Preprint} arXiv:1604.06313 [hep-th])


\bibitem{Durka:2016eun} Durka R 2017 Resonant algebras and gravity 
{\it J.\ Phys.} A {\bf 50} 145202  ({\it Preprint} arXiv:1605.00059 [hep-th])  


\bibitem{Concha:2016tms} Concha P K, Merino N  and Rodr\'iguez  E K 2017 Lovelock gravities from Born-Infeld gravity theory 
{\it Phys.\ Lett.} B {\bf 765} 395-401  ({\it Preprint} arXiv:1606.07083 [hep-th])


\bibitem{Ipinza:2016con} Ipinza M C,  Concha P K,  Ravera L  and Rodr\'iguez E K 2016 On the Supersymmetric Extension of Gauss-Bonnet like Gravity 
{\it J. High Energy Phys.} JHEP09(2016)007 ({\it Preprint} arXiv:1607.00373 [hep-th])


\bibitem{Concha:2016zdb} Concha P K, Fierro O and Rodr\'iguez E K 2016 In\"on\"u-Wigner Contraction and $D=2+1$ Supergravity  
{\it Eur.\ Phys.\ J.} C {\bf 77} 48 ({\it Preprint} arXiv:1611.05018 [hep-th]) 


\bibitem{Penafiel:2016ufo} Pe\~nafiel D M and Ravera L 2017 Generalized In\"on\"u-Wigner Contraction as $S$-Expansion with Infinite Semigroup and Ideal Subtraction 
{\it J.\ Math.\ Phys.}  {\bf 58} 081701   ({\it Preprint} arXiv:1611.05812 [hep-th])


\bibitem{Artebani:2016gwh} Artebani M, Caroca R, Ipinza M C, Pe\~nafiel D M and Salgado P 2016 Geometrical aspects of the Lie algebra S-expansion procedure 
{\it J.\ Math.\ Phys.}  {\bf 57} 023516  ({\it Preprint} arXiv:1602.04525 [math-ph])


\bibitem{Ipinza:2016bfc} Ipinza M C, Lingua F, Pe\~nafiel D M and Ravera L 2016 An Analytic Method for $S$-Expansion involving Resonance and Reduction 
{\it Fortsch.\ Phys.}  {\bf 64} 854-80   ({\it Preprint} arXiv:1609.05042 [hep-th])





%\bibitem{Nesterenko2012} Maryna Nesterenko, ``S-expansions of three-dimensional Lie algebras,'' Institute of Mathematics of NAS of Ukraine, 3 Tereshchenkivs'ka Str., Kyiv-4, 01601 Ukraine, arXiv:1212.1820; Maryna Nesterenko, ``S-expansions of three-dimensional Lie algebras,'' Group analysis of differential equations and integrable systems, 147-154, Department of Mathematics and Statistics, University of Cyprus, Nicosia, 2013.


\bibitem {webJava} https://github.com/SemigroupExp/Sexpansion/releases/tag/v1.0.0

\bibitem{Inostroza:2017ezc} Inostroza C, Kondrashuk I, Merino N and Nadal F 2017 A Java library to perform S-expansions of Lie algebras 
{\it Preprint} arXiv:1703.04036 [cs.MS]

%\bibitem {S10} Andreas Distler, Chris Jefferson, Tom Kelsey and Lars Kotthof, ``The Semigroups of Order 10,'' Proceedings. DOI 10.1007/978-3-642-33558-7\_63

%Andreas Distler, Chris Jefferson, Tom Kelsey and Lars Kotthof, ``The Semigroups of Order 10,'' Principles and Practice of Constraint Programming Volume 7514 of the series Lecture Notes in Computer Science pp 883-899, Springer Berlin Heidelberg. 18th International Conference, CP 2012, Québec City, QC, Canada, October 8-12, 2012. Proceedings. DOI 10.1007/978-3-642-33558-7\_63
% http://link.springer.com/chapter/10.1007%2F978-3-642-33558-7_63


\bibitem{Hildebrant} Hildebrant J 2001 Handbook of Finite Semigroup Programs 
{\it LSU Mathematics Electronic Preprint Series} {\it Preprint} 2001-24




%\bibitem{Nesterenko2012} Nesterenko M 2013  S-expansions of three-dimensional Lie algebras {\it 6th International Workshop Group analysis of differential equations and integrable systems GADEIS-VI: Proceedings of Conference  (Protaras, June 2012)} ed O O Vaneeva, C Sophocleous, R O Popovych, P G L. Leach, V M Boyko and P A Damianou (Nicosia: Department of Mathematics and Statistics, University of Cyprus) pp 147-154 ({\it Preprint} arXiv:1212.1820[math-ph])

%\bibitem {jama} http://math.nist.gov/javanumerics/jama/

%\bibitem {wiki} https://github.com/SemigroupExp/Sexpansion/wiki



%\bibitem{Burde} D. Burde, ``Degenerations of 7-dimensional nilpotent Lie algebras,'' Comm. Algebra {\bf 33} (2005), 1259.

%\bibitem{Popovych} Popovych D.R. and Popovych R.O., ``Lowest dimensional example on non-universality of generalized In\"{o}n\"{u}-Wigner contractions,'' J. Algebra 324 (2010), 2742-2756.

%\bibitem{Arnowitt:1975xg} R.~L.~Arnowitt, P.~Nath and B.~Zumino, ``Superfield Densities and Action Principle in Curved Superspace,'' Phys.\ Lett.\  {\bf 56B}, 81 (1975).
	
%\bibitem{Akulov:1975ax} V.~P.~Akulov, D.~V.~Volkov and V.~A.~Soroka, ``Gauge Fields on Superspaces with Different Holonomy Groups,'' JETP Lett.\  {\bf 22}, 187 (1975) [Pisma Zh.\ Eksp.\ Teor.\ Fiz.\  {\bf 22}, 396 (1975)].


%\bibitem{Dauria} L. Castellani, R. D'Auria and P. Fre, Supergravity and Superstrings: A Geometric Perspective, World Scientific, Singapore, 1991.

%\bibitem{Inostroza} C.~Inostroza, I.~Kondrashuk, N.~Merino, F.~Nadal, ``Algorithm to find S-related Lie algebras,'' Will be submitted soon.

%\bibitem{CS_GR} P.~K.~Concha, C.~Inostroza, N.~Merino and E.~K.~Rodr\'iguez, ``Chern-Simons gravities related with General Relativity,'' Will be submitted soon.

%\bibitem{CS_PL_Gravity} R.~Durka, C.~Inostroza and N.~Merino, ``Pure Lovelock gravity from Chern-Simons theory,'' Work in progress.

\end{thebibliography}
\end{document}